\documentstyle[12pt]{article}

\makeatletter
\def\@sect#1#2#3#4#5#6[#7]#8{\ifnum #2>\c@secnumdepth
    \def\@svsec{}\else
    \refstepcounter{#1}\edef\@svsec{\csname the#1\endcsname.\hskip 1em }\fi
    \@tempskipa #5\relax
    \ifdim \@tempskipa>\z@
    \begingroup #6\relax
    \@hangfrom{\hskip #3\relax\@svsec}{\interlinepenalty \@M #8\par}
    \endgroup
    \csname #1mark\endcsname{#7}\addcontentsline
    {toc}{#1}{\ifnum #2>\c@secnumdepth \else
     \protect\numberline{\csname the#1\endcsname}\fi
           #7}\else
    \def\@svsechd{#6\hskip #3\@svsec #8\csname #1mark\endcsname
          {#7}\addcontentsline
          {toc}{#1}{\ifnum #2>\c@secnumdepth \else
     \protect\numberline{\csname the#1\endcsname}\fi
           #7}}\fi
     \@xsect{#5}}
\def\label#1{\@bsphack\if@filesw {\let\thepage\relax
   \xdef\@gtempa{\write\@auxout{\string
   \newlabel{#1}{{\thesection.\@currentlabel}{\thepage}}}}}\@gtempa
   \if@nobreak \ifvmode\nobreak\fi\fi\fi\@esphack}
\def\@eqnnum{(\thesection.\theequation)}
\def\section{\setcounter{equation}{0} \@startsection {section}{1}{\z@}{-3.5ex
   plus -1ex minus -.2ex}{2.3ex plus .2ex}{\Large\bf}}
\newcount\@minsofar
\newcount\@min
\newcount\@cite@temp
\def\@citex[#1]#2{%
\if@filesw \immediate \write \@auxout {\string \citation {#2}}\fi
\@tempcntb\m@ne \let\@h@ld\relax \def\@citea{}%
\@min\m@ne%
\@cite{%
  \@for \@citeb:=#2\do {\@ifundefined {b@\@citeb}%
    {\@h@ld\@citea\@tempcntb\m@ne{\bf ?}%
    \@warning {Citation `\@citeb ' on page \thepage \space undefined}}%
{\@minsofar\z@ \@for \@scan@cites:=#2\do {%
  \@ifundefined{b@\@scan@cites}%
    {\@cite@temp\m@ne}
    {\@cite@temp\number\csname b@\@scan@cites \endcsname \relax}%
\ifnum\@cite@temp > \@min
    \ifnum\@minsofar = \z@
      \@minsofar\number\@cite@temp
      \edef\@scan@copy{\@scan@cites}\else
    \ifnum\@cite@temp < \@minsofar
      \@minsofar\number\@cite@temp
      \edef\@scan@copy{\@scan@cites}\fi\fi\fi}\@tempcnta\@min
  \ifnum\@minsofar > \z@ 
    \advance\@tempcnta\@ne
    \@min\@minsofar
    \ifnum\@tempcnta=\@minsofar 
      \ifx\@h@ld\relax
        \edef \@h@ld{\@citea\csname b@\@scan@copy\endcsname}%
    \else \edef\@h@ld{\ifmmode{-}\else--\fi\csname b@\@scan@copy\endcsname}%
      \fi
    \else \@h@ld\@citea\csname b@\@scan@copy\endcsname
          \let\@h@ld\relax
  \fi 
\fi}%
\def\@citea{,\penalty\@highpenalty\,}}\@h@ld}{#1}}
\def\appendixname{Appendix}
\def\appendix{\par
  \def\pre@section{\appendixname{}}
  \setcounter{section}{1}
  \@addtoreset{equation}{section}
  \def\thesection{\Alph{section}}
  \def\theequation{\arabic{equation}}}

\newcounter{@sc}
\newcounter{@scp}
\newcounter{@t}
\newlength{\@x}
\newlength{\@xa}
\newlength{\@xb}
\newlength{\@y}
\newlength{\@ya}
\newlength{\@yb}
\newsavebox{\@pt}
\def\bezier#1(#2,#3)(#4,#5)(#6,#7){\c@@sc#1\relax
 \c@@scp\c@@sc \advance\c@@scp\@ne
 \@xb #4\unitlength \advance\@xb -#2\unitlength \multiply\@xb \tw@
 \@xa #6\unitlength \advance\@xa -#2\unitlength
 \advance\@xa -\@xb \divide\@xa\c@@sc
 \@yb #5\unitlength \advance\@yb -#3\unitlength \multiply\@yb \tw@
 \@ya #7\unitlength \advance\@ya -#3\unitlength
 \advance\@ya -\@yb \divide\@ya\c@@sc
 \setbox\@pt\hbox{\vrule height\@halfwidth depth\@halfwidth
 width\@wholewidth}\c@@t\z@
 \put(#2,#3){\@whilenum{\c@@t<\c@@scp}\do
 {\@x\c@@t\@xa \advance\@x\@xb \divide\@x\c@@sc \multiply\@x\c@@t
 \@y\c@@t\@ya \advance\@y\@yb \divide\@y\c@@sc \multiply\@y\c@@t
 \raise \@y \hbox to \z@{\hskip \@x\unhcopy\@pt\hss}\advance\c@@t\@ne}}}

\makeatother
\begin{document}
\addtolength{\unitlength}{-0.5\unitlength}
\def\t{\theta}
\def\ov{\overline}
\def\a{\alpha}
\def\b{\beta}
\def\d{\delta}
\def\g{\gamma}
\def\wt{\widetilde}
\def\m{\overline m}
\def\f{\overline f}
\def\p{\overline p}
\def\n{\overline n}
\def\q{\overline q}
\def\w{\omega}
\def\ds{\displaystyle}
\def\s{\sigma}
\def\G00{\overline\Gamma_0}
\centerline{\bf\LARGE The vertex formulation}
\vspace{0.5cm}
\centerline{\bf\LARGE  of the Bazhanov -- Baxter model}
\vspace{1cm}
\centerline{\large S.M. Sergeev\footnote{E-mail: sergeev\_ms@mx.ihep.su}}
\vspace{0.2cm}
\centerline{ Branch Institute for Nuclear  Physics,}
\centerline{ Institute for High Energy Physics,
Protvino, Moscow Region, Russia}
\vspace{0.5cm}
\centerline{\large
V.V. Mangazeev\footnote{On leave of absence from the Institute for
High Energy Physics, Protvino, Moscow Region, Russia,
E-mail: vvm105@phyvs1.anu.edu.au}}
\vspace{0.2 cm}
\centerline{\small Department of Theoretical Physics, RSPhysSE,}
\centerline{\small
Australian National University, Canberra, ACT 0200, Australia}
\vspace{0.5 cm}
\centerline{\large and }
\vspace{0.5 cm}
\centerline{\large Yu.G. Stroganov\footnote{E-mail: stroganov@mx.ihep.su}}
\centerline{\small Institute for High Energy Physics,}
\centerline{\small Protvino, Moscow Region, Russia}
\vspace{1cm}

\vspace{1cm}
\centerline{Abstract}
In this paper we formulate an integrable model on the simple cubic
lattice. The $N$ -- valued spin variables of the model belong to edges of
the lattice. The Boltzmann weights of the model obey the vertex type
Tetrahedron Equation. In the thermodynamic limit our model is equivalent
to the Bazhanov -- Baxter Model. In the  case when $N=2$ we reproduce the
Korepanov's and Hietarinta's solutions of the Tetrahedron equation as some
special cases.

\newpage

\section{Introduction}

Recently two new solutions of the
vertex type Tetrahedron equation \cite{Tetrah1,Tetrah2,Tetrah3}
for the number of spin states $N=2$ \cite{Korepanov,Hietarinta}
were obtained. In our previous paper we have tried
to generalize these solutions for  $N>2$ and for general spectral
parameters. We have succeeded there in a generalization of the
solution from Ref. \cite{Hietarinta} for arbitrary $N$.

For the case $N=2$ the solution proposed by Hietarinta appears to be
a some special case of the planar limit of the Bazhanov -- Baxter solution
\cite{mss2}. Recall that in the Bazhanov -- Baxter model (BBM) \cite{BB}
$N$ -- valued spin variables belong to the vertices of the
elementary cubes of the lattice and the Boltzmann weights in the
Tetrahedron Equation (TE) are parameterized by the
angles of a tetrahedron \cite{kms}.

As is known the Bazhanov-Baxter model can not be directly
reformulated as a vertex type
model using an obvious duality between vertex and interaction-round-cube
formulations. For example, for the case $N=2$ (Zamolodchikov model)
\cite{Tetrah1} such a duality requires an invariance of the weight
functions with respect to a recolouring of any face of the elementary
cube.  It is known \cite{B} that the Boltzmann weights of
Zamolodchikov model in general do not possess this symmetry
(despite the fact that the absolute values of the Boltzmann weights do).

Nevertheless in the particular limit when all four vertices of the
tetrahedron belong to the same plane (the planar limit), it is possible to
rewrite the Boltzmann weights using 2-state edge variables only and
as a result to
obtain the vertex solution of the TE from Ref. \cite{Hietarinta}.
Note that for $N>2$ the  solution of the TE from Ref. \cite{mss2} does not
coincide with the planar limit of the BBM and seems to be new.

Attempts to remove the planar limit restriction for this solution have
been failed.
Instead  we have obtained a complete (depending on three
arbitrary angles) vertex solution of the TE for general number of
spin variables $N$. This solution at $N=2$ reproduces the solutions of
Korepanov and Hietarinta in the static and planar limits correspondingly.
This new model in the thermodynamic limit coincides with the BBM.
However, due to the vertex form, this formulation may be
useful for a more careful investigation of the model. Namely,
one can try to formulate the Bethe -- ansatz, construct a functional
equation for the transfer matrices analogically to the two -- dimensional
case. Also one can try to construct a three dimensional generalization of
the $L$ operators and etc.

The paper is organized as follows. In the second section we recall
usual notations for the functions on $Z_N$ which will be used for
the constructing of Boltzmann weights.
In the third section we give an explicit form of the vertex weight
function and show an equivalence of our vertex model with the BBM in the
thermodynamic limit. Symmetry properties of the vertex weight  we
list in the fourth section. Also we give some exotic forms of the
gauges and write out the inversion relation for the weight
functions. The case $N=2$ is considered in some special gauge
in the fifth section, where we show the equivalence of our vertex weight
in the static limit with the solution of the TE proposed by Korepanov.
The sixth section is devoted to the sketch of the proof of the TE for the
vertex weight. At last, in Appendix we collect the  most useful
formulae for $\w$ -- hypergeometric seria with $\w$ being a $N$-th
root of unity.

\section{Notations and definitions.}

In this section we give a list of all necessary definitions and notations.

Denote
\begin{equation}
\w^{1/2}=\exp(\pi i/N),\quad N\in Z.                            \label{2.1}
\end{equation}
Let $x$, $y$, $z$ are three homogeneous complex variables
constrained by Fermat equation
\begin{equation}
x^N+y^N=z^N.                                           \label{2.2}
\end{equation}
Hereafter we use a notation $p=(x,y,z)$ anywhere unless it
will lead to misunderstanding.

Now we define a function $w(p|a)$ by recurrence relation:
\begin{equation}
\ds{w(p|a)\over w(p|0)}=\prod_{s=1}^{a}{y\over z-x\w^s}, \label{2.3}
\end{equation}
where $a$ is an element of $Z_N$.

Following Ref. \cite{VVB} we will choose a normalization factor $w(p|0)$
as follows.
First let us set $z=1$ and
consider the case $|x|<1$.
Then we can choose $y$ as
\begin{equation}
y=(1-x^N)^{1/N}.           \label{2.4}
\end{equation}
For such $x$, $y$ and $z$ we put
\begin{equation}
w(p|0)=y^{(1-N)/2}\prod_{j=1}^{N-1}(1-\w^{-j}x)^{j/N}.   \label{2.5}
\end{equation}
With such a normalization the function $w(p|a)$ satisfies
\begin{equation}
\prod_{a=0}^{N-1}w(p|a)=1.         \label{2.6}
\end{equation}

Further we can analytically continue formulas (\ref{2.4}-\ref{2.6})
over $x$ into the whole complex plane with cuts from the points
$x=\w^n$, $n=0,\ldots,N-1$ to infinity.
For such $x$ and $y$ we will say that
the point $p=(x,y,1)$ belongs to the main branch
$\Gamma_0$ of some covering curve
$\Gamma$ on which the function $w(p|a)$ is well defined.
If we go under the cut around the point $x=\w^n$ in the anti-clockwise
direction, then $w(p|0)$ is multiplied on the phase factor $(-1)^{N-1}\w^n$.
Restoring $z$ dependence it is easy to check that
\begin{equation}
w(\w^nx,y,z|m)=w(x,y,z|m+n),\quad m,n\in Z_N,\>p=(x,y,z)\in\Gamma_0.
\label{2.7}
\end{equation}

Now let us consider a region on  $\Gamma_0$ such that
\begin{equation}
-2\pi/N<{\rm Arg}(x/z)<0    \label{2.9}
\end{equation}
and denote it as $\G00$. It is easily to show that for the points
$p\in\G00$ we have
\begin{equation}
-\pi/N<{\rm Arg}(y/z)<\pi/N.            \label{2.9a}
\end{equation}

Then for the given point $p=(x,y,z)\in\G00$ define a new point $Op\in\G00$ as
\begin{equation}
Op=(z,\w^{1/2}y,\w x).             \label{2.10}
\end{equation}

Using these notations
we have the following property of the $w$ function:
\begin{equation}
w(p|a)w(Op|-a)\Phi(a)\exp({i\pi (N^2-1)\over 6N})=1,\quad a\in Z_N,\>
p\in\G00,                    \label{2.11}
\end{equation}
where
\begin{equation}
\Phi(a)=\w^{a(a-N)/2}.                                    \label{2.12}
\end{equation}

In Appendix we will give a set of the most useful formulae and identities for
the $w$ function.

\section{The vertex weights.}

For a given spherical triangle with the angles $\t_1,\t_2,\t_3$
and corresponding linear angles (i.e. sides of the spherical triangle)
$a_1,a_2,a_3$ define four points $p_i\in\G00$:
\begin{eqnarray}
&\ds  x(p_1)=\w^{-1/2}\exp (i{a_3\over N})\sqrt[N]{\sin\b_1\over\sin\b_2},\;
y(p_1) = \exp ( i{\b_1\over N})\sqrt[N]{\sin a_3\over\sin\b_2};&\nonumber\\
&\ds  x(p_2)=\w^{-1/2}\exp (i{a_3\over N})\sqrt[N]{\sin\b_2\over\sin\b_1},\;
y(p_2) = \exp ( i{\b_2\over N})\sqrt[N]{\sin a_3\over\sin\b_1};&\nonumber\\
&\ds  x(p_3)=\w^{-1}\exp (i{a_3\over N})\sqrt[N]{\sin\b_3\over\sin\b_0},\;
y(p_3) = \exp ( -i{\b_3\over N})\sqrt[N]{\sin a_3\over\sin\b_0};&\nonumber\\
&\ds  x(p_4)=\w^{-1}\exp (i{a_3\over N})\sqrt[N]{\sin\b_0\over\sin\b_3},\;
y(p_4) = \exp ( -i{\b_0\over N})\sqrt[N]{\sin a_3\over\sin\b_3};&\nonumber\\
&z(p_i)=1,\,i=1,2,3,4;&                              \label{3.1}
\end{eqnarray}
where $\b_i$ being usual linear excesses
\begin{equation}
\ds\b_0=\pi-{a_1+a_2+a_3\over 2},\;\b_i=\pi-\b_0-a_i.\label{3.2}
\end{equation}
Further we will consider (\ref{3.1}) as
a definition for the points $p_i\equiv p_i(a_1,a_2,a_3)$.

Let $\rho_k$, $k=1,2,3$ be normalization factors:
\begin{equation}
\rho_k={\Biggl(}{\ds\sin a_k\over
\ds2\cos\beta_0/2\cos\beta_1/2\cos\beta_2/2\cos\beta_3/2}
{\Biggr)}^{N-1\over N}.  \label{3.2a}
\end{equation}
With these notations the vertex weight function is
\begin{equation}
\ds R^{j_1,j_2,j_3}_{i_1,i_2,i_3}=
\ds\d_{j_2+j_3,i_2+i_3}\w^{j_3(j_1-i_1)}\rho_3
{w(p_1|i_1-i_2)w(p_2|j_1-j_2)\over
 w(p_3|i_1-j_2)w(p_4|j_1-i_2)}.                       \label{3.3}
\end{equation}
Such weight functions obey the Tetrahedron Equation (see section 6)
and hence define an exactly solvable lattice model.

Now we will show that such a model is equivalent in the thermodynamic limit
to the BBM \cite{BB}. In the BBM to each cube
of the lattice we assign a weight function depending on eight corner
spins (see Fig. 1).

\begin{picture}(600,265)
\put(0,50){
\begin{picture}(500,200)
\multiput(140,0)(120,0){2}{\line(0,1){120}}
\multiput(140,0)(0,120){2}{\line(1,0){120}}
\multiput(140,0)(0,120){2}{\line(-1,1){60}}
\put(80,180){\line(1,0){120}}\put(80,180){\line(0,-1){120}}
\put(200,180){\line(1,-1){60}}
\multiput(200,180)(0,-20){6}{\line(0,-1){12}}
\multiput(80,60)(20,0){6}{\line(1,0){12}}
\multiput(255,5)(-30,30){2}{\line(-1,1){20}}
\multiput(140,0)(120,0){2}{\circle*{15}}
\multiput(140,120)(120,0){2}{\circle*{15}}
\multiput(80,60)(120,0){2}{\circle*{15}}
\multiput(80,180)(120,0){2}{\circle*{15}}
\put(300,80){\large $=\quad W(a|e,f,g|b,c,d|h)$}
\put(150,10){$e$}\put(270,10){$d$}\put(150,100){$a$}
\put(270,100){$f$}\put(212,186){$b$}\put(92,190){$g$}
\put(92,65){$c$}\put(212,65){$h$}
\end{picture}
}
\put(320,0){\Large\bf Fig. 1}
\end{picture}

\noindent
We choose the weight function in the following form
\begin{eqnarray}
&\ds W(a|e,f,g|b,c,d|h|a^B_1,a^B_2,a^B_3) = &\nonumber\\
&\rho_3\ds \sum_{\s}
{w(Op^B_4|d-h-\s)w(Op^B_3|a-g-\s)\over
w(Op^B_2|e-c-\s)w(Op^B_1|f-b-\s)}
\w^{\s(h+g-c-b)},&                                      \label{3.4}
\end{eqnarray}
where  $p^B_i=p_i(a^B_1,a^B_2,a^B_3)$.
Here $a_i^B$ are the sides of the spherical triangle
in notations of Ref. \cite{BB,kms}.
In fact, expression (\ref{3.4}) coincides with the weight function
from Ref. \cite{kms} up to some gauge and normalization multipliers.

Consider now a chain of $n$ weights (\ref{3.4}) in the direction $a-g$
with the cyclic boundary conditions. This chain defines a
weight function for some two -- dimensional model, closely connected with
the homogenious generalized  Chiral Potts model \cite{BB,CPM}.
The weight function of this two -- dimensional model looks like
\begin{eqnarray}
&\ds {\cal W}(A,B,C,D|a^B_1,a^B_2,a^B_3)=
\prod_{\a} W(a_\a|c_\a,b_\a,a_{\a+1}|b_{\a+1},c_{\a+1},d_\a|d_{\a+1})=
&\nonumber\\
&\ds \sum_{\{\s_\a\}}\prod_{\a}\>\rho_3\>
{w(Op^B_4|\hat d_\a-\s_\a)w(Op^B_3|\hat a_\a-\s_\a)\over
w(Op^B_2|\hat c_\a-\s_\a)w(Op^B_1|\hat b_\a-\s_\a)}
\w^{\s_\a(d_{\a+1}+a_{\a+1}-c_{\a+1}-b_{\a+1})},
&                                                         \label{3.5}
\end{eqnarray}
where the capital spin $A = \{a_\a\}$ and $\hat a_\a\equiv a_\a-a_{\a+1}$,
etc (see Fig. 2).

\begin{picture}(600,265)
\put(0,50){
\begin{picture}(500,200)
\multiput(140,0)(120,0){2}{\line(0,1){120}}
\multiput(140,0)(0,120){2}{\line(1,0){120}}
\multiput(140,0)(0,120){2}{\line(-1,1){60}}
\put(80,180){\line(1,0){120}}\put(80,180){\line(0,-1){120}}
\put(200,180){\line(1,-1){60}}
\multiput(200,180)(0,-20){6}{\line(0,-1){12}}
\multiput(80,60)(20,0){6}{\line(1,0){12}}
\multiput(255,5)(-30,30){2}{\line(-1,1){20}}
\multiput(140,0)(120,0){2}{\circle*{15}}
\multiput(140,120)(120,0){2}{\circle*{15}}
\multiput(80,60)(120,0){2}{\circle*{15}}
\multiput(80,180)(120,0){2}{\circle*{15}}
\put(320,90){$\rightarrow$}
\multiput(400,30)(120,0){2}{\circle*{25}}
\multiput(400,150)(120,0){2}{\circle*{25}}
\multiput(460,90)(120,0){2}{\circle*{25}}
\put(400,30){\line(1,1){120}}
\put(400,150){\line(1,-1){120}}
\put(520,150){\line(1,-1){60}}
\put(520,30){\line(1,1){60}}
\put(415,160){$A$}\put(535,160){$B$}\put(415,10){$C$}\put(535,10){$D$}
\put(450,110){$\Sigma$}\put(570,110){$\Sigma'$}
\put(150,10){$c_\a$}\put(270,10){$d_\a$}\put(150,100){$a_\a$}
\put(270,100){$b_\a$}\put(212,186){$b_{\a+1}$}\put(92,190){$a_{\a+1}$}
\put(92,65){$c_{\a+1}$}\put(212,65){$d_{\a+1}$}
\end{picture}
}
\put(320,0){\Large\bf Fig. 2}
\end{picture}

Move now our frame of view to the right at the one half step of the two --
dimensional lattice (see the right part of the Fig. 2).
The points $A$ and $C$ disappear from our frame, but there will
appear the right neighbour $\Sigma'$
 of our previous spin of the summation $\Sigma$.
Then we get the four -- spin weight $\cal S$:
\begin{eqnarray}
&\ds S(\Sigma,B,\Sigma',D|a^B_1,a^B_2,a^B_3)=&\nonumber\\
&\ds =\prod_{\a}\>\rho_3\>
{w(Op^B_4|\hat d_\a-\s_\a)w(Op^B_3|\hat b_\a-\s'_\a)\over
w(Op^B_2|\hat d_\a-\s'_\a)w(Op^B_1|\hat b_\a-\s_\a)}
\w^{(\s_\a-\s'_\a)(d_{\a+1}-b_{\a+1})}\equiv&\nonumber\\
&\ds\equiv\prod_{\a} R^{\ds -\s_\a,-\hat d_\a,b_{\a+1}-d_{\a+1}}
_{\ds -\s'_\a,-\hat b_\a,\phantom{+1}b_{\a}-d_{\a}}
(a^B_2,\pi-a^B_1,\pi-a^B_3),&                               \label{3.6}
\end{eqnarray}
where the vertex weight $R$ is defined by (\ref{3.3}).
The last expression  in (\ref{3.6}) differs
from the two -- dimensional projection of the vertex lattice by a slight
modification of boundary conditions. So
the BBM with the weight functions (\ref{3.4}), depending on
$a^B_i$, is equivalent to the vertex model with the weight functions
(\ref{3.3}), depending on the $a_i$ such as
\begin{equation}
a_1=a^B_2,\;a_2=\pi-a^B_1,\;a_3=\pi-a^B_3                   \label{3.7}
\end{equation}
in the thermodynamic limit.

\section{Symmetry properties.}

The weight function of the BBM (\ref{3.4})
is symmetric with respect to the cube symmetry group up to some
multiplicative gauge transformations. In the case of the vertex weight
function (\ref{3.3}) corresponding gauge transformations are the Fourier
ones. To simplify all formulas we will use  convenient
operator notations. We will consider our weight (\ref{3.3})
as an operator acting in the tensor product of three linear $N$ --
dimensional spaces so that
\begin{equation}
R^{j_1,j_2,j_3}_{i_1,i_2,i_3}=<i_1,i_2,i_3|R|j_1,j_2,j_3>.
\label{4.1}
\end{equation}
Define operators of the Fourier transformation and of the spin inversion:
\begin{equation}
<i|F|j>={1\over\sqrt{N}}\w^{ij},\;<i|J|j>=\d_{i,-j}.\label{4.2}
\end{equation}
Then the inversion relation for the vertex weight is
\begin{eqnarray}
&\ds R(a_1,a_2,a_3)\, J\otimes J\otimes J\, R(-a_1,-a_2,-a_3)\, J\otimes
J\otimes J\, =&\nonumber\\
&\ds = 1\otimes 1\otimes 1 \,
\Phi(a_1,a_2,a_3),&       \label{4.3}
\end{eqnarray}
where
\begin{equation}
\Phi(a_1,a_2,a_3)=\Biggl({\ds\sin\beta_0/2\over
\ds\cos\beta_1/2\cos\beta_2/2\cos\beta_3/2}\Biggr)^{2{N-1\over N}}.
\label{4.3a}
\end{equation}
In fact, expression (\ref{4.3a}) coincides explicitly
with the inversion factor for BBM \cite{VVB}.

To write down the symmetry properties of the weight (\ref{3.3})
we need to define the permutation operators $P_{ij}$
\begin{equation}
P_{12}|i_1,i_2,i_3>=|i_2,i_1,i_3>                          \label{4.4}
\end{equation}
and similarly for $P_{13}$, $P_{23}$.

Let $t_1, t_2, t_3$ be transpositions in the corresponding vector spaces
and the total transposition sign $t$:
\begin{equation}
R^t(a_1,a_2,a_3)=R^{t_1t_2t_3}(a_1,a_2,a_3).     \label{4.5}
\end{equation}
Then the crossing relation is
\begin{equation}
\ds 1\otimes 1\otimes J\> R^{t_1t_2}(a_1,a_2,a_3)\>
1\otimes 1\otimes J = R(\pi-a_1,\pi-a_2,a_3)    \label{4,6}
\end{equation}
and all five space permutations are given by
\begin{eqnarray}
&F^{-1}\otimes F^{-1}\otimes F^{-1} R(a_1,a_2,a_3) F\otimes F\otimes F =
P_{13} R^t(a_3,a_2,a_1) P_{13},&\nonumber\\
&J\otimes J\otimes F R(a_1,a_2,a_3) J\otimes J\otimes F^{-1} =
P_{12} R^t(a_2,a_1,a_3) P_{12}.&\nonumber\\
&F\otimes 1\otimes 1 R(a_1,a_2,a_3) F^{-1}\otimes 1\otimes 1 =
P_{23} R^t(a_1,a_3,a_2) P_{23},\label{4.7}\\
&F^{-1}\otimes F^{-1}\otimes J R(a_1,a_2,a_3)F\otimes F\otimes J =
P_{23} P_{12} R(a_3,a_1,a_2)P_{12} P_{23},&\nonumber\\
&J\otimes F\otimes F R(a_1,a_2,a_3) J\otimes F^{-1}\otimes F^{-1} =
P_{12} P_{23} R(a_2,a_3,a_1)P_{23} P_{12}.&\nonumber
\end{eqnarray}

Combining Fourier transformations with diagonal gauge transformations,
one can obtain another forms of the $R$ -- matrix.
In general these combined transformations are the gauge
transformations of the lattice, but not the gauge transformations
of the Tetrahedron Equation.

Note that there exists the conservation law $i_2+i_3=j_2+j_3$ for
the weight (\ref{3.3}).
Below we write out some combined transformations which lead to another
forms of the spin conservations laws:
\begin{eqnarray}
&\ds N^{-1}\sum_{\a_3,\b_3}\w^{\ds \a_3j_3-\b_3i_3}
\left({\Phi(\a_3)\over\Phi(\b_3)}\right)^{\epsilon}
R^{j_1,j_2,\a_3}_{i_1,i_2,\b_3}(a_1,a_2,a_3)=&\nonumber\\
&\ds =\d (j_3+j_1-i_3-i_1-\epsilon (j_2-i_2))\Phi(j_2-i_2)^{-\epsilon}
\w^{\ds -i_3(j_2-i_2)}\times&\nonumber\\
&\times
\rho_3\>\ds {w(p_1(a_1,a_2,a_3)|i_1-i_2)w(p_2(a_1,a_2,a_3)|j_1-j_2)\over
      w(p_3(a_1,a_2,a_3)|i_1-j_2)w(p_4(a_1,a_2,a_3)|j_1-i_2)},&\label{4.9}
\end{eqnarray}
\begin{eqnarray}
&\ds N^{-2}\sum_{\a_m,\b_m}
\w^{\ds \a_1j_1-\b_1i_1+\a_2j_2-\b_2i_2}
\left({\Phi(\a_2)\over\Phi(\b_2)}\right)^{\epsilon}
R^{\a_1,\a_2,j_3}_{\b_1,\b_2,i_3}(a_1,a_2,a_3)=&\nonumber\\
&\ds =\d (j_2+j_1-i_2-i_1-\epsilon (j_3-i_3))\Phi(j_3-i_3)^{\epsilon}
\w^{\ds j_2(i_3-j_3)}\times&\nonumber\\
&\times
\ds \rho_2\>{w(p_1(a_1,a_3,a_2)|-j_1-j_3)w(p_2(a_1,a_3,a_2)|-i_1-i_3)\over
      w(p_3(a_1,a_3,a_2)|-j_1-i_3)w(p_4(a_1,a_3,a_2)|-i_1-j_3)},&\label{4.10}
\end{eqnarray}
\begin{eqnarray}
&\ds N^{-3}\sum_{\a_m,\b_m}
(\prod_m\w^{\ds \a_mj_m-\b_mi_m})
\left({\Phi(\a_1)\over\Phi(\b_1)}\right)^{\epsilon}
R^{\a_1,\a_2,\a_3}_{\b_1,\b_2,\b_3}(a_1,a_2,a_3)=&\nonumber\\
&\ds =\d (j_1+j_2-i_1-i_2-\epsilon (j_3-i_3))\Phi(j_3-i_3)^{-\epsilon}
\w^{\ds i_1(i_3-j_3)}\times&\nonumber\\
&\times
\ds \rho_1\>{w(p_1(a_3,a_2,a_1)|j_3-j_2)w(p_2(a_3,a_2,a_1)|i_3-i_2)\over
      w(p_3(a_3,a_2,a_1)|j_3-i_2)w(p_4(a_3,a_2,a_1)|i_3-j_2)},&\label{4.11}
\end{eqnarray}
where $\d(a)=\d_{a,0}$,
$\Phi(a)$ and $\rho_k$ are defined by (\ref{2.12}, \ref{3.2a})
 and $\epsilon$ is an
arbitrary integer. Another choices of
the diagonal $\Phi$ factors lead us to complicated nonmultiplicative
expressions for the weights. The exception
is the case $N=2$.

\section{The case $N=2$}

In this section we consider the case $N=2$ in a special gauge in which
our $R$ matrix (\ref{3.3}) can be reduced to the vertex solutions
of Korepanov and Hietarinta in static and planar limits correspondingly.

For the case $N=2$ the list of suitable Fourier transformations
enlarges. Namely,
\begin{eqnarray}
&\ds N^{-3}\xi\sum_{\a_m,\b_m}
\left(\prod_{m=1}^3\w^{\ds
\a_mj_m-\b_mi_m}{\Phi(\b_m)\over\Phi(\a_m)}\right)
R^{\a_1,\a_2,\a_3}_{\b_1,\b_2,\b_3}(a_1,a_2,a_3)=&\nonumber\\
&\ds =\d(i_1+j_1+i_3+j_3-i_2-j_2)
\w^{\ds (i_3-j_3)(i_2-j_3)}\times&\nonumber\\
&\ds\times {w(r_1|-i_1+j_2-j_3)w(r_3|i_1-i_2+j_3)\over
      w(Or_0|i_1-i_2+i_3)w(Or_2|-i_1+j_2-i_3)}\equiv
\overline R^{j_1,j_2,j_3}_{i_1,i_2,i_3},&                      \label{5.1}
\end{eqnarray}
where $\w=-1$,
\begin{equation}
\xi=\left({4\sqrt{\cot\b_0/2\dots\cot\b_3/2}\,}\right)^{N-1\over N},
                                                                \label{5.2}
\end{equation}
and four points $r_i$ are given by
\begin{eqnarray}
&\ds x(r_i)=\exp (-i\b_i/N),\;y(r_i)=\w^{1/4}\sqrt[N]{2\sin\b_i},&
\nonumber\\
&\ds z(r_i)=\exp (i\b_i/N),\;i=0,1,2,3.&                       \label{5.3}
\end{eqnarray}
Due to the total symmetry, this transformation is the
gauge transformation of the Tetrahedron Equation, so this weight
$\overline R$ obeys the Tetrahedron Equation (see the section 6).

When $N=2$ the function $w$ is very simple:
\begin{equation}
\ds {w(r_i|1)\over w(r_i|0)}=
\exp(i{\pi\over 4})\sqrt{\tan{\b_i\over 2}}.            \label{5.4}
\end{equation}
Define
\begin{equation}
\ds\sqrt{\tan{\b_i\over 2}}=t_i,\;i=0,1,2,3.                     \label{5.5}
\end{equation}
Then we can represent
the weights (\ref{5.1}) by the following compact table:

\vspace*{0.5cm}

\begin{tabular}{|ccccrcrcl|}\hline
&&&&&&&& \\
$\ds\ov R^{0,0,0}_{0,0,0}$&=&$\ds\ov R^{0,1,1}_{0,1,1}$&=&
$\ds\ov R^{1,0,1}_{1,0,1}$&=&$\ds\ov R^{1,1,0}_{1,1,0}$ &=&
$\ds 1$ \\
&&&&&&&& \\
$\ds\ov R^{1,1,1}_{1,1,1}$&=&$\ds\ov R^{1,0,0}_{1,0,0}$&=&
$\ds\ov R^{0,1,0}_{0,1,0}$&=&$\ds\ov R^{0,0,1}_{0,0,1}$ &=&
$\ds t_0t_1t_2t_3$ \\
&&&&&&&& \\
\hline
&&&&&&&& \\
$\ds\ov R^{0,1,0}_{0,0,1}$&=&$\ds\ov R^{0,0,1}_{0,1,0}$&=&
$\ds -\ov R^{1,0,0}_{1,1,1}$&=&$\ds -\ov R^{1,1,1}_{1,0,0}$ &=&
$\ds  t_0t_1$ \\
&&&&&&&& \\
$\ds\ov R^{1,0,1}_{1,1,0}$&=&$\ds\ov R^{1,1,0}_{1,0,1}$&=&
$\ds -\ov R^{0,1,1}_{0,0,0}$&=&$\ds -\ov R^{0,0,0}_{0,1,1}$ &=&
$\ds t_2t_3$ \\
&&&&&&&& \\
\hline
&&&&&&&& \\
$\ds\ov R^{1,1,1}_{0,1,0}$&=&$\ds\ov R^{1,0,0}_{0,0,1}$&=&
$\ds -\ov R^{0,0,1}_{1,0,0}$&=&$\ds -\ov R^{0,1,0}_{1,1,1}$ &=&
$\ds it_0t_2$ \\
&&&&&&&& \\
$\ds\ov R^{0,0,0}_{1,0,1}$&=&$\ds\ov R^{0,1,1}_{1,1,0}$&=&
$\ds -\ov R^{1,1,0}_{0,1,1}$&=&$\ds -\ov R^{1,0,1}_{0,0,0}$ &=&
$\ds  -it_1t_3$ \\
&&&&&&&& \\
\hline
&&&&&&&& \\
$\ds\ov R^{0,0,1}_{1,1,1}$&=&$\ds\ov R^{1,1,1}_{0,0,1}$&=&
$\ds\ov R^{1,0,0}_{0,1,0}$&=&$\ds\ov R^{0,1,0}_{1,0,0}$ &=&
$\ds t_0t_3$ \\
&&&&&&&& \\
$\ds\ov R^{1,1,0}_{0,0,0}$&=&$\ds\ov R^{0,0,0}_{1,1,0}$&=&
$\ds\ov R^{0,1,1}_{1,0,1}$&=&$\ds\ov R^{1,0,1}_{0,1,1}$ &=&
$\ds t_1t_2$ \\
&&&&&&&& \\
\hline
\end{tabular}

\vspace*{0.5cm}

The weight, defined by this table, differs from
(\ref{5.1}) by some normalization factor.
Using the following property of a spherical triangle:
\begin{equation}
\ds\tan{\b_i\over 2}\tan{\b_j\over 2}=
\tan{\a_k\over 2}\tan{\a_l\over 2},\; \{i,j,k,l\}=\{0,1,2,3\},\label{5.6}
\end{equation}
where $\a_i$ are the angle excesses of the spherical triangle,
we can easily obtain the static limit of the $\ov R$
 (the case when $\a_0=0$). This static limit
appears to be the solution of the Tetrahedron Equation
proposed by Korepanov \cite{Korepanov}.
Moreover, in the planar limit when $\b_2=0$ the
vertex weight (\ref{5.1}) coincides with the $N=2$ solution by Hietarinta
\cite{Hietarinta,mss2}.

\section{The Tetrahedron Equation.}

The vertex form of the TE is the following:
\begin{eqnarray}
&{\ds\sum_{k_1,k_2,k_3,\atop k_4,k_5,k_6}}
R^{k_1,k_2,k_3}_{i_1,i_2,i_3}
R'^{j_1k_4k_5}_{\phantom{,}k_1i_4i_5}
R''^{j_2j_4k_6}_{\phantom{,,}k_2k_4i_6}
R'''^{j_3j_5j_6}_{\phantom{,,,}k_3k_5k_6}
=&\nonumber\\
&={\ds\sum_{k_1,k_2,k_3,\atop k_4,k_5,k_6}}
R'''^{k_3,k_5,k_6}_{\phantom{,,,}i_3,i_5,i_6}
R''^{k_2k_4j_6}_{\phantom{,,}i_2i_4k_6}
R'^{k_1j_4j_5}_{\phantom{,}i_1k_4k_5}
R^{j_1j_2j_3}_{k_1k_2k_3}
&.\label{6.1}
\end{eqnarray}
A complete solution of this equation is parameterized by six angles of a
tetrahedron (five of them are independent):
\begin{eqnarray}
&\ds R = R(\t_1,\t_2,\t_3),&\nonumber\\
&\ds R' = R(\t_1,\t_4,\t_5),&\nonumber\\
&\ds R'' = R(\pi-\t_2,\t_4,\t_6)&\nonumber\\
&\ds R''' = R(\t_3,\pi-\t_5,\t_6).&                        \label{6.2}
\end{eqnarray}
The ordering of the dihedral angles is natural with respect to
the numbering of the spaces and differs from that in the
standard equation (2.2) in Ref. \cite{B}.

For each vertex in (\ref{6.1}) let $a_i$ be the corresponding planar angles:
\begin{eqnarray}
&\ds (\t_1,\t_2,\t_3)\rightarrow (a_1,a_2,a_3),&\nonumber\\
&\ds (\t_1,\t_4,\t_5)\rightarrow (a'_1,a'_2,a'_3),&\nonumber\\
&\ds (\pi-\t_2,\t_4,\t_6)\rightarrow (a''_1,a''_2,a''_3),&\nonumber\\
&\ds (\t_3,\pi-\t_5,\t_6)\rightarrow (a'''_1,a'''_2,a'''_3),&\label{6.3}
\end{eqnarray}
Then planar angles of four weights are constrained as follows
\begin{eqnarray}
&a_3''=a_3'-a_3,\quad a_1'''=a_1''-a_1',&\nonumber\\
&a_2'''=a_2''-a_1,\quad a_3'''=a_2'-a_2.&\label{6.3a}
\end{eqnarray}

Four vertex weights (\ref{3.3}) with the angles defined by (\ref{6.3})
and satisfying to (\ref{6.3a})
obey  Tetrahedron Equation (\ref{6.1}).
In this section we give a sketch of the proof
of this statement.

Let us
substitute (\ref{3.3}) in (\ref{6.1}). Due to the spin conservation laws,
six summations in the both sides reduce to three ones. It is useful
to choose the indices $k_1,k_2,k_4$ as  independent spins of the
summations. The summands in the left and right hand sides are the products
of the phase factors $\w^{\dots}$ and $w$ functions. First let us
collect in both sides all
factors depending on the spin $k_1$. The sums over $k_1$ have the form
$_2\Psi_2$ (see Appendix). As the first step let us apply to these $_2\Psi_2$
$(\tau\rho)^2$ transformations (see formula (\ref{a9}) from  Appendix).
As a result there appear extra $w$ functions, depending on $k_2-k_4$,
and we
demand a cancellation of these extra factors with similar
$w$ functions in the left and right hand sides. It gives us some
algebraic constraints on parameters of weights.

Further summations over $k_2$ and $k_4$ become independent.
Moreover, there are no phase factors, depending on $k_2$ and $k_4$,
and we can sum over $k_2$ and $k_4$ using the ``Star -- Square''
relation (see  formula (\ref{a14}) from Appendix). Finally  sums
over $k_1$ are both of the type $_4\Psi_4$ and have a similar spin
structure. Imposing necessary constraints among parameters of
weights we come to the equality of the left
and right hand sides of the TE.

We will not write out here algebraic constraints coming from
the cancellation
of all $w$ functions, depending on $k_2-k_4$, the ``Star -- Square''
applicability conditions (\ref{a15}) and the
coincidence of the final expressions.
All calculations are direct but rather tedious.

As a result we obtain that all restrictions on parameters of
weight functions are satisfied automatically if we take into account
parameterization (\ref{3.1}) and
constraints between angles (\ref{6.3}-\ref{6.3a}).

\section{Acknoledgements}
We would like to thank R. Baxter for
reading the manuscript and valuable comments, and
V.V. Bazhanov for useful discussions and suggestions.
This research has been partially supported by National Science Foundation
Grant PHY -- 93 -- 07 -- 816 and by International Science Foundation
(ISF), Grant RMM000.
\appendix
\section*{Appendix}

In this Appendix we collect  useful formulae in the theory
of $\w$ -- hypergeometric seria with $\w$ being a $N$-th root
of unity. In fact, these formulae (or their particular cases)
appeared in many papers devoted to the Chiral Potts Model
\cite{OCPM1,OCPM2,OCPM3} and to the TE.
Let us define $_r\Psi_r$ series as
\begin{equation}
\ds_r\Psi_r
\left(\left.{(p_1,m_1)\dots (p_r,m_r)\atop (p'_1,m'_1)\dots (p'_r,m'_r)}
\right|n\right)
=\sum_{\s\in Z_N}
{w(p_1|m_1+\s)\dots w(p_r|m_r+\s)\over
w(p'_1|m'_1+\s)\dots w(p'_r|m'_r+\s)}{\w^{\ds n\s}\over\sqrt{N}}.\label{a1}
\end{equation}

Discuss now a role of the normalization.
Spin independent factors in all identities in this Appendix are given
for the case when all arguments of the $w$ functions in the left and right
hand sides belong to the region $\G00$ (see (\ref{2.9})).
If we abandon  restriction (\ref{2.9}) then the
phases of $w$'s can be chosen in such a way that the corresponding formulae
will still remain correct.

To simplify all notations
we will omit arguments in the components of points $p_i$ and imply that
\begin{equation}
\ds p_i = (x_i,y_i,z_i)\label{a2}
\end{equation}
for every $i$. In last formulas of this Appendix we will use also
points $q_i$. In this case we will explicitly
point out by upper subscript the corresponding point
\begin{equation}
\ds q_i=(x^q_i,y^q_i,z^q_i).              \label{a2a}
\end{equation}

There will appear many new points on the Fermat curve
in right hand sides of formulae. In these
cases we have to introduce new letters for $y$ components. They have to be
defined by (\ref{2.4}) (and belong to region (\ref{2.9a}) in the accordance
with our agreement).

All formulae in this Appendix are formulae of the summation
(they exist for $r=1,2,3$) and  symmetry transformation
formulae (they exist for $r=1,2,3,4$).

We begin with a cyclic analog of Ramanujan summation formula for $r=1$.
In fact, this is nothing else as restricted Star-Triangle relation
of Bazhanov-Baxter model (see Ref. \cite{RSTR})

\begin{eqnarray}
&\ds _1\Psi_1\left(\left.{(p_1,m_1)\atop (p_2,m_2)}\right|n\right)=&
\label{a3}\\
&\ds =\Phi_0
\left({\xi\over y_1y_2}\right)^{N-1\over 2}
\ds {w(z_1y_2,\xi,y_1z_2|-n)w(x_1z_2,\xi,\w z_1x_2|m_1-m_2)
\over \w^{\ds nm_2}\> w(x_1y_2,\xi,\w y_1x_2|m_1-m_2-n)}=&\nonumber\\
&\ds=
\left({\w^{-1/2}\xi\over y_1y_2}\right)^{N-1\over 2}\hspace{-0.1cm}
\ds{\ds \Phi_0^{-1}\>
\w^{\ds -nm_1}\> w(y_1x_2,\w^{-1/2}\xi,x_1y_2|m_2-m_1+n)\over
w(y_1z_2/\w,\w^{-1/2}\xi,z_1y_2|n)
w(z_1x_2,\w^{-1/2}\xi,x_1z_2|m_2-m_1)},&\nonumber
\end{eqnarray}
\noindent
\vspace*{0.1cm}
where

\begin{equation}
\Phi_0=\exp\Bigl\{i\pi{\ds (N-1)(N-2)\over \ds 12N}\Bigr\}.     \label{a3a}
\end{equation}
\vspace*{0.1cm}

For the proof of this formula see, for example, Ref. \cite{RSTR,StSq}.

Here we  give also two forms of inversion relations for $w$ functions
which have been used for the proof
of inversion relation
(\ref{4.3}):
\begin{equation}
\sum_\s{w(\w x,\w y,z|m_1+\s)\over w(x,y,z|m_2+\s)}=
N\d_{\ds m_1,m_2}\left({\w^{1/2}x\over y}\right)^{N-1}.\label{a41}
\end{equation}
\vspace{0.1cm}
\begin{equation}
\sum_\s{w(x,y,z|m_1+\s)\over w(x,y,\w z|m_2+\s)}=
N\d_{\ds m_1,m_2}\left({\w^{1/2}x\over y}\right)^{N-1}.\label{a42}
\end{equation}
\vspace*{0.2cm}
\noindent Note that  points
$(\w x,\w y, z)$ and $(x, y,\w z)$ do not belong to
the region $\G00$ and we define for these cases
\begin{equation}
{\ds w(\w x,\w y, z|0)\over \ds w(x,y,z|0)}=
-\w^{1/2}{\ds y\over \ds z-\w x}               \label{a43}
\end{equation}
and
\begin{equation}
{\ds w(x,y,\w z|0)\over \ds w(x,y,z|0)}=
-\w^{1/2}{\ds z-x\over \ds y},               \label{a44}
\end{equation}
where $(x,y,z)\in \G00$.

To obtain symmetry formulae for higher $r$, we use the following simple fact.
Let $g_1$ and $g_2$ be arbitrary functions on $Z_N$. If
\begin{equation}
\tilde g_i(k) = \sum_{\s\in Z_N}g_i(\s){\w^{\ds k\s}\over\sqrt{N}},
\label{a5}
\end{equation}
then
\begin{equation}
\sum_{\s\in Z_N}g_1(\s)g_2(\s) = \sum_{\s\in Z_N}\tilde g_1(\s)
\tilde g_2(-\s).\label{a6}
\end{equation}
Using this, it is easily to obtain the following symmetry transformation
for $_2\Psi_2$:
\begin{eqnarray}
&\ds _2\Psi_2\left(\left.{(p_1,m_1)(p_3,m_3)\atop
(p_2,m_2)(p_4,m_4)}\right|n\right)=
\phantom{a}_2\Psi_2\left(\left.{(q_1,0)\atop (q_2,n)}{(q_3,m_4-m_3+n)\atop
(q_4,m_1-m_2)}\right|m_2-m_3\right) &\nonumber\\
&\ds\times\left({\xi_{12}\xi_{43}\over y_1y_2y_3y_4}\right)^{N-1\over2}
\w^{\ds -nm_3}{w(x_1z_2,\xi_{12},\w z_1x_2|m_1-m_2)\over
w(z_3x_4,\xi_{43},x_3z_4|m_4-m_3)},&\label{a7}
\end{eqnarray}
where
\begin{eqnarray}
q_1=(z_1y_2,\xi_{12},y_1z_2),&& q_3=(y_3x_4,\xi_{43},x_3y_4),\nonumber\\
q_2=(y_3z_4/\w,\xi_{43},z_3y_4),&&
q_4=(x_1y_2,\xi_{12},\w y_1x_2).\label{a8}
\end{eqnarray}
This relation has appeared originally as the $(\tau\rho)$ transformation in
Ref. \cite{StSq} for the BB weight function. Note that
$(\tau\rho)^6=1$. In this paper we have used $(\tau\rho)^2$ transformation:
\begin{eqnarray}
&\ds _2\Psi_2\left(\left.{(p_1,m_1)(p_3,m_3)\atop
(p_2,m_2)(p_4,m_4)}\right|n\right)=&\nonumber\\
&\ds=\phantom{a}_2\Psi_2\left(\left.{(s_1,0)\atop (s_2,m_2-m_3)}
{(s_3,m_1-m_4-n)\atop
(s_4,-n)}\right|m_3-m_4\right)\times &\nonumber\\
&\ds\times\> \w^{\ds -nm_2-(m_2-m_3)(m_4-m_3)}\left({\Lambda\Lambda'\over
\xi_{12}\xi_{43} y_1y_2y_3y_4}\right)^{N-1\over2}\times
 & \nonumber\\
&\ds\times {w(x_1z_2,\xi_{12},\w z_1x_2|m_1-m_2)\over
w(z_3x_4,\xi_{43},x_3z_4|m_4-m_3)}
{w(z_1z_3y_2y_4,\Lambda,z_2z_4y_1y_3|-n)\over
w(x_1x_3y_2y_4,\Lambda',\w x_2x_4y_1y_3|-\n)}
,&\label{a9}
\end{eqnarray}
where
\begin{equation}
\ds \n=n-m_1-m_3+m_2+m_4,\label{a10}
\end{equation}
and
\begin{eqnarray}
s_1=(y_1z_2\xi_{43},\Lambda,z_3y_4\xi_{12}),&&
s_3=(x_1y_2\xi_{43},\Lambda',y_3x_4\xi_{12}),\nonumber\\
s_2=(y_1x_2\xi_{43},\Lambda',x_3y_4\xi_{12}),&&
s_4=(z_1y_2\xi_{43},\Lambda, y_3z_4\xi_{12}).\label{a11}
\end{eqnarray}
The list of the symmetry formulae for $_2\Psi_2$ we finish by the
$T=(\tau\rho)^3$:
\begin{eqnarray}
&\ds _2\Psi_2\left(\left.{(p_1,m_1)(p_3,m_3)\atop
(p_2,m_2)(p_4,m_4)}\right|n\right)=
\phantom{a}_2\Psi_2\left(\left.{(\p_1,-m_3)(\p_3,-m_1)\atop
(\p_2,-m_4)(\p_4,-m_2)}\right|\n\right)\times &\nonumber\\
&\ds\times\>\w^{\ds -nm_2-m_4\n}\left({\xi_{12}\xi_{43}\xi_{32}\xi_{41}\over
\Lambda\Lambda'}\right)^{N-1\over2}
{w(x_1z_2,\xi_{12},\w z_1x_2|m_1-m_2)\over
w(z_3x_4,\xi_{43},x_3z_4|m_4-m_3)}\times &\nonumber\\
&\ds\times{w(z_1z_3y_2y_4,\Lambda,z_2z_4y_1y_3|-n)\over
w(x_1x_3y_2y_4,\Lambda',\w x_2x_4y_1y_3|-\n)}
\ds {w(x_3z_2,\xi_{32},\w z_3x_2|m_3-m_2)\over
w(z_1x_4,\xi_{41},x_1z_4|m_4-m_1)},&\label{a12}
\end{eqnarray}
where
\begin{eqnarray}
\p_1=(z_3\Lambda',\xi_{32}\xi_{43}y_1,x_3\Lambda),&&
\p_3=(z_1\Lambda',\xi_{41}\xi_{12}y_3,x_1\Lambda),\nonumber\\
\p_2=(z_4\Lambda',\w\xi_{41}\xi_{43}y_2,\w x_4\Lambda),&&
\p_4=(z_2\Lambda',\xi_{32}\xi_{12}y_4,\w x_2\Lambda).\label{a13}
\end{eqnarray}
Note that in the case when ${x_1x_3\over x_2x_4} = \w {z_1z_3\over
z_2z_4}$, (\ref{a12}) becomes the Star -- Star relation for the BBM, and
$p_i=\p_i$.

To obtain a summation formula for $_2\Psi_2$, consider (\ref{a7})
and set $n=0$ and $q_1=q_2$. Then applying (\ref{a3}) to the
right hand side of (\ref{a7}), we obtain ``Star -- Square''
relation:
\begin{eqnarray}
&\ds _2\Psi_2\left(\left.{(p_1,m_1)(p_3,m_3)\atop
(p_2,m_2)(p_4,m_4)}\right|0\right)=
\left({\w^{1/2}\Lambda'\over y_1y_2y_3y_4}\right)^{N-1\over2}
\w^{\ds -(m_2-m_3)(m_1-m_2)}\times &\nonumber\\
&\ds\times\>\Phi_0\>
w(x_2x_4y_1y_3,\w^{-1/2}\Lambda',x_1x_3y_2y_4|m_2+m_4-m_1-m_3)\times
&\nonumber\\
&\ds\times {w(x_1z_2,\xi_{12},\w z_1x_2|m_1-m_2)\over
w(z_3x_4,\xi_{43},x_3z_4|m_4-m_3)}
{w(x_3z_2,\xi_{32},\w z_3x_2|m_3-m_2)\over
w(z_1x_4,\xi_{41},x_1z_4|m_4-m_1)},&\label{a14}
\end{eqnarray}
where the parameters in the left hand side have to obey a special
restriction:
\begin{equation}\label{a15}
\ds {y_1y_3\over y_2y_4}=\w {z_1z_3\over z_2z_4},
\end{equation}
and the phases in the right hand side are constrained by
\begin{equation}\label{a15.1}
\ds {\xi_{12}\over\xi_{43}}={y_1z_2\over y_4z_3},\;\;\;
{\xi_{32}\over\xi_{41}}={y_3z_2\over y_4z_1},\;\;\;
\Lambda'=\w^{-1/2}\xi_{12}\xi_{32}{y_4\over z_2}.
\end{equation}
Further we will try to avoid such long notations as in (\ref{a12})
and (\ref{a14}).
Extra $w$ multipliers in all consequent formulae will have the same
structure as in the right hand side of (\ref{a12}) and so we will use
only $\xi_{ij}$ to denote a whole argument dependence of $w$.

Consider now $r=3$. A summation formula can be obtained summing
(\ref{a12}) over $n$ with the help of the restricted Star-Triangle
relation (\ref{a3}).
The result reads
\begin{eqnarray}\label{a16}
&\ds _3\Psi_3\left(\left.{(p_1,m_1)(p_3,m_3)(q_1,m_2+m_4-\lambda)\atop
(p_2,m_2)(p_4,m_4)(q_2,m_1+m_3-\lambda)}\right|0\right)=&\nonumber\\
&\ds=\Phi_0^{-1}\left({\xi_{12}\xi_{43}\xi_{32}\xi_{41}\over
y_1y_2y_3y_4\Xi}\right)^{N-1\over2}\times&\nonumber\\
&\times\ds{\ds\w^{\ds (m_4-\lambda)(m_1+m_3-m_2-m_4)}\over\ds
w(x_1x_3z_2z_4,\Xi,\w^2x_2x_4z_1z_3|m_1+m_3-m_2-m_4)}\times&\nonumber\\
&\times\ds{w(\xi_{12}|m_1-m_2)w(\xi_{32}|m_3-m_2)\over
w(\xi_{43}|m_4-m_3)w(\xi_{41}|m_4-m_1)}
{w(\p_1|\lambda-m_3)w(\p_3|\lambda-m_1)\over
w(\p_2|\lambda-m_4)w(\p_4|\lambda-m_2)},&
\end{eqnarray}
where $w(\xi_{ij})$ and $w(\p_i)$ are the same as in (\ref{a12}) and
\begin{eqnarray}
&\ds q_1=(\w x_2x_4\Lambda,y_2y_4\Xi,z_2z_4\Lambda'),&\nonumber\\
&\ds q_2=(x_1x_3\Lambda, y_1y_3\Xi,\w z_1z_3\Lambda').&\label{a17}
\end{eqnarray}
Note, that the formula (\ref{a16}) is symmetric with respect to any
permutation of $p_1,p_3,q_1$ and $p_2,p_4,q_2$.
The Star -- Triangle relation \cite{OCPM2,OCPM3}
for the Chiral Potts Model is a special
case of (\ref{a16}).

To obtain symmetry relations for $r=3$ and $r=4$, we have to use
(\ref{a6}), apply (\ref{a3}) or $T$ transformation correspondingly,
cancel extra $w$ factors (this gives some constraints) and then,
using (\ref{a6}) again, obtain the corresponding $_r\Psi_r$ in the
right hand side. The formula for $r=3$ reads
\begin{eqnarray}\label{a18}
&\ds _3\Psi_3\left(\left.{(p_1,m_1)(p_3,m_3)(q_1,l_1)\atop
(p_2,m_2)(p_4,m_4)(q_2,l_2)}\right|0\right)=&\nonumber\\
&\ds =\left({\xi_{12}\xi_{43}\xi_{32}\xi_{41}\xi\over
\Lambda'^2\Lambda y^q_1y^q_2}\right)^{N-1\over2}
\w^{\ds (l_1-m_2)(m_1+m_3-m_2-m_4)}\times &\nonumber\\
&\times\ds{w(\xi_{12}|m_1-m_2)w(\xi_{32}|m_3-m_2)\over
w(\xi_{43}|m_4-m_3)w(\xi_{41}|m_4-m_1)}
{w(\xi|l_2+m_2+m_4-l_1-m_1-m_3)\over
w(\lambda|l_2-l_1)}\times&\nonumber\\
&\times\ds _3\Psi_3\left(\left.{(\p_1,-m_3)(\p_3,-m_1)(\q_1,l_1-m_2-m_4)\atop
(\p_2,-m_4)(\p_4,-m_2)(\q_2,l_2-m_1-m_3)}\right|0\right),&
\end{eqnarray}
where the connections between arguments in the left hand side are
\begin{equation}
\ds {y^p_1y^p_3y^q_1\over y^p_2y^p_4y^q_2}=\w{z^p_1z^p_3z^q_1\over
z^p_2z^p_4z^q_2},
\qquad
{\Lambda\over\lambda}={z^p_2z^p_4y^p_1y^p_3\over z^q_1y^q_2}\label{a19}
\end{equation}
and new arguments in the right hand side of (\ref{a18}) are
\begin{eqnarray}
&\ds (\xi)=(\w x^p_2x^p_4x^q_2y^p_1y^p_3y^q_1,\xi,
\w x^p_1x^p_3x^q_1y^p_2y^p_4y^q_2),&
\nonumber\\
&\ds (\lambda)=(z^q_1x^q_2,\lambda,x^q_1z^q_2),&\nonumber\\
&\ds \q_1=(x^q_1y^q_2\Lambda',\xi,\w\lambda x^p_2x^p_4y^p_1y^p_3),&\nonumber\\
&\ds \q_2=(y^q_1x^q_2\Lambda',\xi,\w\lambda x^p_1x^p_3y^p_2y^p_4).&\label{a20}
\end{eqnarray}
Note that this formula is a symmetry transformation for something.
Denote (\ref{a18}) as $\rho_3$. Let $\tau_3$ be a permutation
transformation,
reordering the columns in $_3\Psi_3$ as $\tau_3(1,2,3)=(2,3,1)$.
Then $(\tau_3\rho_3)^6=1$.

The last formula is a symmetry transformation for $_4\Psi_4$.
A derivation of it is described before formula (\ref{a18}).
Let the structure of a set $q_i,\q_i,\chi_{ij},\Delta,\Delta'$ is defined
identically to that of $p_i,\p_i,\xi_{ij},\Lambda,\Lambda'$. Then
\begin{eqnarray}\label{a21}
&\ds _4\Psi_4\left(\left.{(p_1,m_1)(p_3,m_3)(q_1,l_1)(q_3,l_3)\atop
(p_2,m_2)(p_4,m_4)(q_2,l_2)(q_4,l_4)}\right|0\right)=&\\
&\ds=\left({\xi_{12}\xi_{43}\xi_{32}\xi_{41}
\chi_{12}\chi_{43}\chi_{32}\chi_{41}\over
\Lambda'\Lambda\Delta'\Delta}\right)^{N-1\over2}
{\w^{(l_2-m_2)(m_1+m_3-m_2-m_4)}\over\Phi(m_1+m_3-m_2-m_4)}
\times &\nonumber\\
&\times \ds{w(\xi_{12}|m_1-m_2)w(\xi_{32}|m_3-m_2)\over
w(\xi_{43}|m_4-m_3)w(\xi_{41}|m_4-m_1)}
{w(\chi_{12}|l_1-l_2)w(\chi_{32}|l_3-l_2)\over
w(\chi_{43}|l_4-l_3)w(\chi_{41}|l_4-l_1)}\times &\nonumber\\
&\times \ds _4\Psi_4\left(\left.{(\p_1,-m_3)(\p_3,-m_1)
(\q_1,l_1-m_2-m_4)(\q_3,l_3-m_2-m_4)\atop
(\p_2,-m_4)(\p_4,-m_2)(\q_2,l_2-m_1-m_3)(\q_4,l_4-m_1-m_3)}\right|0\right),&
\nonumber
\end{eqnarray}
where the constraints are
\begin{equation}
\ds {y^p_1y^p_3y^q_1y^q_3\over y^p_2y^p_4y^q_2y^q_4}\ds =
\w{z^p_1z^p_3z^q_1z^q_3\over z^p_2z^p_4z^q_2z^q_4}
\ds = \w^{-1}{x^p_1x^p_3x^q_1x^q_3\over x^p_2x^p_4x^q_2x^q_4}
\label{a22}
\end{equation}
and
\begin{equation}
\ds {\Lambda\over\Delta}=\w^{-1/2}
{y^p_1y^p_3z^p_2z^p_4\over z^q_1z^q_3y^q_2y^q_4},\quad
\ds {\Lambda'\over\Delta'}=\w^{1/2}
{y^p_1y^p_3x^p_2x^p_4\over x^q_1x^q_3y^q_2y^q_4},\label{a23}
\end{equation}
and the spins in (\ref{a21}) are not independent:
\begin{equation}
\ds m_1+m_3+l_1+l_3 = m_2+m_4+l_2+l_4.
\end{equation}

\end{document}